\documentstyle[prl,aps,epsf,floats,twocolumn]{revtex}

\newcommand{\vk}{{\vec{k}}}

\newcommand{\vs}{{\vec \sigma}}
\newcommand{\vS}{{\vec S}}

\def\lst2{{(l^*)^2}}

\newcommand{\eeq}{\end{equation}}

\newcommand{\beq}{\begin{equation}}

\def\t{{\theta}}

\def\half{{1\over2}}

\def\bea{\begin{eqnarray}}
\def\eea{\end{eqnarray}}
\def\a{{\alpha}}
\def\b{{\beta}}
\def\g{{\gamma}}
\def\d{{\delta}}

\def\e{{\epsilon}}

\def\prl{{Phys. Rev. Lett.\ }}
\def\prb{{Phys. Rev. B\ }}
\begin{document}
\draft
\flushbottom
\twocolumn[
\hsize\textwidth\columnwidth\hsize\csname @twocolumnfalse\endcsname
\title{Interactions and Disorder in Quantum Dots: Instabilities and Phase Transitions}
\author{Ganpathy Murthy$^1$ and Harsh Mathur$^2$}
\address{$^1$Department of Physics and Astronomy, University of Kentucky, Lexington KY 40506-0055
\\$^2$Department of Physics, Case Western Reserve University, Cleveland OH}
\date{\today}
\maketitle

\begin{abstract}
Using a fermionic renormalization group approach we analyse a model
where the electrons diffusing on a quantum dot interact via
Fermi-liquid interactions. Describing the single-particle states by
Random Matrix Theory, we find that interactions can induce phase
transitions (or crossovers for finite systems) to regimes where
fluctuations and collective effects dominate at low
energies. Implications for experiments and numerical work on quantum
dots are discussed.
\end{abstract}
]

A variety of finite many-body physical systems are so complex that one
can only approach them statistically. Random Matrix Theory
(RMT)\cite{mehta} has emerged as a unifying language to describe
nucleii, atoms, and molecules\cite{guhr} at intermediate energies. In
these systems, the low-energy physics is well described in terms of
single-particle levels forming shells, but {\it many-body} states at
high enough excitation energy are randomized by
interactions\cite{2brim}, and well described by RMT. More recently,
RMT has been applied to quantum chaos and mesoscopic physics,
including quantum dots\cite{guhr,review}. Here, in contrast to earlier
applications, RMT is a valid description of {\it single-particle}
states, with the external disorder potential now inducing the
randomization between these states. What happens when disorder and
interactions are both present?  This question is relevant to the
tunnelling of electrons through realistic quantum dots, which has been
the focus of much recent experimental\cite{sivan,expt} and
theoretical\cite{sivan,review,exact,hf,models} interest. In this
Letter we show, using renormalization group (RG) methods, that the
introduction of interactions into quantum dots can produce
phase transitions (actually sharp crossovers for finite system
size) in the limit of weak disorder, leading to behavior qualitatively
different from the noninteracting case.

Let us focus on a two-dimensional quantum dot (QD) of linear size
$L$. The average single-particle level spacing is $\Delta$, the mean
free path is $l_{mf}$, and the diffusion constant is $D=v_F l_{mf}/2$,
where $v_F=k_F/m^*$ is the Fermi velocity, $k_F$ is the Fermi
momentum, $m^*$ is the effective mass of the electron, and $\hbar$ has
been set to 1. The diffusion constant is related to another important
energy scale, the Thouless energy $E_T=D/L^2$, which is the inverse of
the diffusion time through the QD. When the QD is strongly coupled to
leads the incoming electron's energy becomes uncertain upto
$E_T$, and therefore the electron ``samples'' $g=E_T/\Delta$
single-particle states\cite{review}, leading to a tunnelling
conductance of $ge^2/h$.

We will be interested in QDs with weak disorder ($g\gg1$), very weakly
coupled to the leads. In this case\cite{alt1} within any energy band
of width $E_T$ (the Thouless band) the energy levels are coupled by
correlations described by RMT. These correlations make the energies
rigid so that the probability of finding two close levels vanishes as
$|\e_1-\e_2|^{\beta}$. Depending on the symmetries of the problem,
there are ten universal random matrix ensembles\cite{mehta,zirnbauer}.
We focus on the orthogonal ensemble with spin degeneracy for which
$\beta=1$. Experiments\cite{sivan,expt} can indirectly
measure the energy to add an electron to the dot (called
$\Delta_2$). For the case with spin degeneracy the distribution
of $ \Delta_2 $ is the sum of the distribution of single particle
spacings (given approximately by the famous Wigner surmise)
and a delta function at zero energy, to account for an opposite
spin electron going into a singly occupied state, leading to the
prediction
\beq
p_{RMT}(\Delta_2)=\half [ \delta(\Delta_2)+
\frac{\pi}{2} \Delta_2 \exp{( -\pi 
\Delta^2_2)} ]
\eeq

This distribution is bimodal and has a peak at zero energy. What is
observed in experiments\cite{sivan,expt} has no bimodality, and looks
more like a symmetric gaussian around some nonzero average. These
features are also observed in numerics\cite{sivan,exact,hf} which makes
it clear that the change of shape
and shift of the distribution is the result of electron-electron
interactions.

The simplest extension of the noninteracting model is the constant
exchange and interaction (CEI) model, where the direct interaction is
modelled by a capacitive term ${\hat Q}^2/2C$, and the exchange by a
$-J{\vec S}^2$ term in the Hamiltonian. The CEI model (or ``universal
Hamiltonian''\cite{models,kurland,brouwer} $H_U$) is a very natural
starting point in the $g\to\infty$ limit. Using $ H_U $ does improve
comparison with experiments\cite{baranger}, but the bimodality remains
at very low temperatures. This does not agree with numerics which
are carried out at $T=0$ and do not show bimodality. 

Motivated by these questions, we have carried out a systematic study
of how interactions affect low-energy properties in a disordered
QD. We use a fermionic RG method developed by one of us\cite{rg-us},
based on the formalism of Shankar\cite{rg-shankar}, in which one
integrates out high-energy states and looks at the flow of couplings
in the low-energy theory. A divergent flow signals a transition to a
phase whose low-energy physics is not perturbatively connected to the
noninteracting limit. For a finite QD the phase transition will be
replaced by a sharp crossover. The advantage of RG over perturbative
methods is that the presence of regimes of qualitatively different
low-energy behavior is made manifest.

We will assume that the interactions are strong compared to the
disorder. The renormalization of the interactions proceeds as in the
clean limit\cite{rg-shankar}, leading to a Fermi liquid (FL) theory
near the Fermi surface. For spinless fermions the FL is charaterized
by a single-particle energy $\e_0(\vk)\approx v_F|\vk-\vk_F|$, and an
interaction function $f(\vk,\vk')$ which parameterizes how the energy
of a quasiparticle (QP) of momentum $\vk$ depends on the occupation of
other QPs
\beq
\e(\vk)=\e_0(\vk)+\sum\limits_{\vk'} f(\vk,\vk')\delta n(\vk')
\eeq
with $\delta n(\vk)$ denoting the change of occupation of state $\vk$
with respect to the filled Fermi sea, which is the ground state. The
low-energy effective FL Hamiltonian is
\beq
H_{FL}=\sum\limits_{\vk} \e_0(\vk) {\hat n}(\vk)+
\half\sum\limits_{\vk\vk'} f(\vk,\vk') 
:{\hat n}(\vk){\hat n}(\vk'):
\eeq
where ${\hat n}(\vk)=c^{\dagger}(\vk)c(\vk)$, and the creation and
annihilation operators obey canonical anticommutation relations.  We
will focus on the Thouless band centered on the Fermi surface. In
particular, there are $g$ momemtum states in this
band which are completely and randomly mixed\cite{alt1} to produce
the eigenstates of the single-particle Hamiltonian (including
disorder) $\phi_\a(\vk)$, with energies $\e_\a$. In terms of these
states  the disordered FL Hamiltonian is
\beq
H_{DFL}=\sum \e_\a \psi_\a^{\dagger}\psi_\a +\half \sum V_{\a\b\g\d} \psi_\a^{\dagger} \psi_\b^{\dagger} \psi_\g\psi_\d
\eeq
We de-dimensionalize the FL interaction by
$f(\vk,\vk')=E_Tu(\t,\t')/g$. Note that $u$ depends only on the
angular position. One resolves the interaction into Fourier
components
\beq
u(\t,\t')=u(\t-\t')=u_0+\sum\limits_{m=1}^{\infty} u_m\cos(m(\t-\t'))
\eeq
We will be interested in the case with spin rotation invariance, where
there are two FL functions, $f_s$ for the singlet channel, and $f_t$
for the triplet channel. The corresponding (anti)symmetrized matrix
elements are
\bea
V^{(s,t)}_{\a\b\g\d}=&{\Delta\over 4}\sum\limits_{\vk,\vk'} u_{s,t}(\t,\t') 
(\phi_\a^*(\vk)\phi_\b^*(\vk')\pm\phi_\a^*(\vk')\phi_\b^*(\vk))\nonumber\\
&(\phi_\d(\vk)\phi_\g(\vk')\pm\phi_\d(\vk')\phi_\g(\vk))
\eea
where the $+$ sign is to be taken with the singlet and the $-$ with
the triplet. Parameterizing the FL interaction as\cite{agd} $
f(\vk,\vk',\vs,\vs')=\Phi(\vk,\vk')+\vs\cdot\vs' Z(\vk,\vk')$, where
the spin of the electron is $\vS=\half\vs$, the singlet and triplet
interactions can be expressed\cite{agd} as $u_{s}=\Phi-3Z$ and
$u_{t}=\Phi+Z$.

In RMT, wavefunctions at different energies are uncorrelated (except
for orthogonality). Denoting an ensemble average 
by $<>$ we have, for the orthogonal ensemble 
\beq
<\phi_\a^*(\vk)\phi_\b(\vk')>=<\phi_\a(-\vk)\phi_\b(\vk')>={\d_{\a\b}\d_{\vk\vk'}\over g}
\eeq
The only matrix elements with nonzero average are 
$<V^{(s)}_{\a\b\b\a}>=u_{s0}\Delta$, $<V^{(t)}_{\a\b\b\a}>=u_{t0}\Delta$, and 
$<V^{(s)}_{\a\a\b\b}>= u_{s}(\pi)\Delta/g$. 
The first two terms when rearranged produce the CEI
model\cite{models}, while the last term is a Cooper interaction, which
however, is suppressed for our Fermi liquid by $1/g$. Since $g$ is
assumed large we ignore it henceforth\cite{models}.

One can similarly calculate the variances of the matrix elements. Here
one has to pay attention to the subtle correlations between different
eigenfunctions induced by orthogonality\cite{mehta}. One finds that
$u_{s0}$ and $u_{t0}$ produce no fluctuations. The general expression is
\beq
<V_{\a\b\g\d}^2>-<V_{\a\b\g\d}>^2={\Delta^2\over 4g^2}\sum\limits_{m=1}^{\infty}u_m^2
\label{variances}\eeq
Kurland {\it et al} argued\cite{kurland} that since these fluctuations
vanish in the limit $g\to\infty$ the universal Hamiltonian is exact in
this limit. On the other hand, because of their large number,
fluctuation terms can have important effects\cite{jacquod}. The best
way to see whether the fluctuation terms are important for low-energy
physics (such as the conductance peak spacing experiments) is to carry
out an RG analysis and see how these terms scale. If interactions
beyond the universal Hamiltonian grow, they will dominate the
low-energy physics {\it no matter how weak they are in the microscopic
Hamiltonian}.

We begin with $g$ states in the Thouless band around the Fermi
surface. We successively integrate out single-particle states farthest
from the Fermi surface, and at some stage in the procedure we have
$g'$ states. We will define the flow parameter $l$ of the RG by
$g'(l)=ge^{-l}$ (with ${d\over dl}=-g'{d\over dg'}$). We do not
rescale the energy, since that would also rescale $\Delta$, which we
want to retain as a physical parameter. We want to integrate out the
states at $\pm g'/2$ to 1-loop order. However, all 1-loop diagrams have
two internal lines. One of these lines is the state at $\pm g'/2$. The
other should be summed over all {\it lower energy} states
$|g''|\le|g'|/2$. This abolishes all
reference to the states $g'/2$ and higher in the effective
theory. Each set of internal lines contributes the following
change to the interaction matrix elements
\bea
&{d V_{\a\b\g\d}^{(s)}\over dl}=-{g'\over2}G_1(\mu,\nu) V_{\a\b\mu\nu}^{(s)}V_{\nu\mu\g\d}^{(s)}+g'G_2(\mu,\nu)\times\nonumber \\
&\bigg(V_{\a\mu\nu\g}^{(s)}V_{\b\nu\mu\d}^{(s)}-{3\over4}V_{\a\mu\nu\g}^{(s+t)}
V_{\b\nu\mu\d}^{(s+t)}+ (\a\leftrightarrow\b)\bigg)\nonumber\\
&{d V_{\a\b\g\d}^{(t)}\over dl}=-{g'\over2}G_1(\mu,\nu) V_{\a\b\mu\nu}^{(t)}V_{\nu\mu\g\d}^{(t)}+g'G_2(\mu,\nu)\nonumber \\
&\bigg(V_{\a\mu\nu\g}^{(t)}V_{\b\nu\mu\d}^{(t)}+{1\over4} V_{\a\mu\nu\g}^{(s+t)}
V_{\b\nu\mu\d}^{(s+t)}- (\a\leftrightarrow\b)\bigg)
\eea
where $V^{(s+t)}=V^{(s)}+V^{(t)}$, 
$G_1(\mu,\nu)=(1-n_{F\mu}-n_{F\nu})/(\e_\mu+\e_\nu)$
and
$G_2(\mu,\nu)=(n_{F\nu}-n_{F\mu})/(\e_\mu-\e_\nu)$. 

In order to make further analytical progress we ensemble average the
internal lines of the diagram $(\mu,\nu)$. The rationale is that in
the course of integrating out states we will have to sum over many
many states $(\mu,\nu)$.  In order to carry out the average we use
\beq
<\phi_\mu^*(1)\phi_\mu(2)\phi_\nu^*(3)\phi_\nu(4)>={\d_{12}\d_{34}\over g^2}-
{\d_{1,-3}\d_{2,-4}+\d_{14}\d_{23}\over g^3}
\eeq
At any stage in the RG it is $g$, and not $g'$, that appears
in these averages, since at one-loop there is no wavefunction
renormalization, and the original wavefunction correlations remain
unchanged. With this averaging, the change in
interaction matrix elements can be absorbed into a change in the FL
parameters. The resulting flow equations to
zeroth order in $1/g$ (for $n\ne 0$) are
\bea
&{du_{sn}(l)\over dl}=Ce^{-l} \bigg((u_{sn}(l))^2-{3\over4}(u_{sn}(l)+u_{tn}(l))^2\bigg)\nonumber\\
&{du_{tn}(l)\over dl}=-Ce^{-l} \bigg((u_{tn}(l))^2+{1\over4} (u_{sn}(l)+u_{tn}(l))^2\bigg)
\eea
where $C=\log{(2)}$ comes from the integration over energies less than
$g'$.  Each moment $n$ of the Fermi liquid function is not mixed with
any other moment $n'$, and $u_{s0},\ u_{t0}$ do not flow. The flow
equations have an explicit dependence on the flow parameter $l$, but
this dependence can be removed by defining a new coupling ${\tilde
u}_n(l)=e^{-l} u_n(l)$, which are the FL parameters for a ``rescaled''
QD with a rescaled Thouless energy $E_T'=e^{-l}E_T$ (corresponding to
$g'=e^{-l}g$ completely randomized states). Rewriting the above
equations in terms of ${\tilde\Phi}_n=({\tilde u}_{sn}+3{\tilde
u}_{tn})/4$ and ${\tilde Z}_n=({\tilde u}_{tn}-{\tilde u}_{sn})/4$ we
obtain
\bea
&{d{\tilde\Phi}_n(l)\over dl}=-{\tilde\Phi}_n-2C{\tilde\Phi}_n^2\nonumber\\
&{d{\tilde Z}_n(l)\over dl}=-{\tilde Z}_n-2C{\tilde Z}_n^2
\eea
The decoupling of ${\tilde\Phi}_n$ and ${\tilde Z}_n$ is a consequence
of spin rotation invariance. From these flow equations we now
construct the phase diagram that is the central result of our paper.
Clearly, there are unstable fixed points at ${\tilde\Phi}_n=-1/2C$ and
${\tilde Z}_n=-1/2C$ which separate weak and strong coupling phases.
\begin{figure}
\narrowtext
\epsfxsize=2.4in\epsfysize=2.4in
\hskip 0.3in\epsfbox{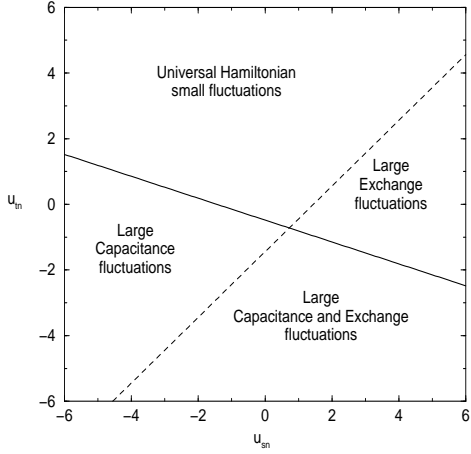}
\vskip 0.15in
\caption{Phase diagram in the $g\to\infty$ limit. Four different phases emerge. 
\label{fig1}}
\end{figure}
There are four phases (Figure 1): (i) A phase where both $\Phi_n$ and
$Z_n$ have small fluctuations, whose low-energy physics is controlled
by the universal Hamiltonian. (ii) A phase where $Z_n$ diverges but
$\Phi_n$ remains small, in which the capacitance will have small
fluctuations but the exchange interaction will acquire large
fluctuations. (iii) A phase where $\Phi_n$ diverges but $Z_n$ remains
small, in which there are large capacitance fluctuations, but the
exchange energy has small fluctuations. (iv) Finally, a phase in which
both $\Phi_n$ and $Z_n$ flow to strong coupling.

The same flow equations and phase diagram are obtained in the unitary
case if time-reversal symmetry is broken by the coupling of
an applied magnetic field to the orbital sector only, as happens in
GaAs dots.

Let us consider the implications for a real quantum dot. Here
$g\approx 10-100$ is typical, which is not particularly large. This
means that the RG flow should be stopped when $g'=ge^{-l}\approx
1$. In the strongly fluctuating phases the fluctuations will increase,
but the flow will be cutoff at the nonuniversal scale $l^*\approx
\log{(g)}$ and the putative phase transition will be replaced by a crossover. 
Nevertheless, we can expect that fluctuations of the
matrix elements play an important role at low energies.

The peak spacing distribution can be qualitatively understood by a
two-level example, considered by Levit and Orgad\cite{hf}. The
resulting peak spacing distribution is approximately gaussian with a
mean position controlled by the constant interaction (which is
$u_{s0},u_{t0}$ in our language), and a variance given approximately
by\cite{hf}
\beq
<\Delta_2^2>=4 \left[ <V_{1221}^2> - < V_{1221} >^2 \right]
+{4\Delta^2\over \pi}
\eeq
If bare interaction matrix elements are used in the above equation,
their contribution is down by $1/g^2$ compared to the single-particle
contribution and negligible (assuming that $|u_n(l=0)|\ll g$).  In the
regime of weak fluctuations this remains true, and the width of the
distribution is controlled mainly by single-particle effects.
However, it is clear that in the regime of strong fluctuations the
variance of the interaction matrix elements can dominate the
single-particle contribution, and thus the width of this distribution
should increase with interaction strength, as has been observed in
numerical work\cite{sivan,exact,hf}.

The flow equations also have implications for interaction effects on
the width of the quasiparticle levels\cite{fock-loc}. Based on bare
matrix elements, the prediction is that quasiparticle peaks should be
relatively well-defined for energies $\e\le\Delta \sqrt{g}$, while
they acquire a width larger than the single particle spacing at larger
energies.  If one is well into a strongly fluctuating phase, one
expects that this threshold energy is pushed down and may even become
zero. In this case the system would not be a Fermi liquid and the
conductance peaks will have a finite width even at $T=0$ (which has
been seen experimentally\cite{abusch}). 

Consider now FL instabilities. In the clean system, the FL becomes
unstable whenever any of the $\Phi_n,\ Z_n\le-1$\cite{agd}. In the
two-level example, the energy to add a particle is controlled by
$V_{\a\b\b\a}$ where one of the levels is occupied and the other
empty. In turn this matrix element is the sum of its average
(controlled by $u_{s0},\ u_{t0}$) and a fluctuation part (controlled
by $u_{sn},\ u_{tn}$). If the fluctuations overpower the constant
interaction, this matrix element can become negative, leading to a
``bunching'' instability\cite{bunch-expt,bunch-th} in which more than
one particle can be added to the QD.  The charge/spin density wave
instability of the clean system when $\Phi_n,\ Z_n\le-1$ will be
modified by disorder and finite-size effects. The characterization of
such a ground state is an open problem.

Our approach can be easily generalized to other
ensembles\cite{mehta,zirnbauer} and the results will be reported in
forthcoming work. It is tempting to speculate about the connection
with a putative metal-insulator transition in an infinite
two-dimensional system\cite{2dmit,belitz}. However, our analysis is
limited to the zero-dimensional case.  Strong fluctuations in the
matrix elements would have implications for the distribution of
persistent currents in mesoscopic structures\cite{persist}. Also
relevant is the effect of a weak magnetic field, which enhances
fluctuations\cite{adam}, on the phase diagram. Finally, for realistic
QD's, corrections to universal RMT behavior (of order
$1/g$)\cite{review} may be important.

In closing, RMT is a rich field with applications to nucleii, atoms
and molecules, where interactions randomize excited
states\cite{2brim}. We have shown that in QD's interactions and
disorder can combine to produce a state in which
fluctuations in interaction matrix elements become large, in which
quasiparticles may cease to be well-defined at arbitrarily low
energies, and the behavior is dominated by collective effects.

We gratefully acknowledge partial support from the NSF from grants
DMR- 98-04983 (HM) and DMR-0071611 (GM), and helpful conversations
with R. Shankar.

 \end{document}